\documentstyle[11pt]{article}
\begin{document}
\title{ Pair production of black holes on cosmic strings}
\author{S.W. Hawking$^a$ and Simon F. Ross$^b$ \\ Department of
  Applied Mathematics and Theoretical Physics \\ University of
  Cambridge, Silver St., Cambridge CB3 9EW \\ $^a$ {\it
    hawking@damtp.cam.ac.uk} \\ $^b${\it S.F.Ross@damtp.cam.ac.uk}}
\date{\today \\ DAMTP/R-95/30}
\maketitle

\begin{abstract}
  We discuss the pair creation of black holes by the breaking of a
  cosmic string. We obtain an instanton describing this process from
  the $C$ metric, and calculate its probability. This is very low for
  the strings that have been suggested for galaxy formation.
\end{abstract}

\pagebreak

The study of black hole pair creation has offered a number of exciting
insights into the nature of quantum gravity, including some further
evidence that the exponential of the black hole entropy really
corresponds to the number of quantum states of the black hole
\cite{ggs,dgkt,dggh,entar}. Black hole pair production is a tunnelling
process, so it can be studied by finding a suitable instanton, that
is, a Euclidean solution which interpolates between the states before
and after the pair of black holes are created. The amplitude for pair
creation is then given by $e^{-I_i}$, where $I_i$ is the action of the
instanton. Black hole pair creation has been commonly studied in the
context of the Ernst metric \cite{Ernst,ggs}, which describes the
creation of a pair of charged black holes by a background electromagnetic
field. The Lorentzian section of the Ernst metric represents a pair of
charged black holes being uniformly accelerated by a background
electromagnetic field.

If we consider the Ernst metric with zero background field, we obtain
a simpler solution called the $C$ metric \cite{Cmetric}. The
Lorentzian section still describes a pair of black holes uniformly
accelerating away from each other, but there is now no background
field to provide the acceleration. This means that there is either a
conical deficit extending from each black hole to infinity, or a
conical surplus running between the two black holes. These can be
thought of respectively as ``strings'' pulling the two black holes
apart, or a ``rod'' pushing them apart.

The purpose of this letter is to argue that the $C$ metric can also be
interpreted as representing pair creation. Specifically, we can
imagine replacing the conical deficit in the $C$ metric with a cosmic
string \cite{ruth}.  The Lorentzian section would then be interpreted
as representing a pair of black holes at the ends of two pieces of
cosmic string, being accelerated away from each other by the string
tension. The Euclidean section of the $C$ metric thus gives an
instanton describing the breaking of a cosmic string, with a pair of
black holes being produced at the terminal points of the string. The
infinite acceleration, zero black hole mass limit of this breaking has
been previously considered in \cite{penrose}.  We will calculate the
action of the Euclidean $C$ metric relative to flat space with a
conical deficit, which gives the approximate rate for cosmic strings
to break by this process.  A similar calculation has previously been
done for the breaking of a string with monopoles produced on the free
ends \cite{vil}, and we will show that our results agree with those,
in the appropriate limit.

The charged $C$ metric solution is
\begin{equation} \label{Cmetric}
  ds^2 = A^{-2} (x-y)^{-2} \left[ G(y) dt^2 - G^{-1}(y) dy^2 +
  G^{-1}(x) dx^2 + G(x) d\varphi^2 \right],
\end{equation}
where
\begin{equation}
  G(\xi) = (1-r_- A \xi) (1 - \xi^2 - r_+ A \xi^3)
\end{equation}
while the gauge potential is
\begin{equation}
  A_\varphi = q (x - \xi_3),
\end{equation}
where $q^2 = r_+ r_-$, and we define $m = (r_+ + r_-)/2$. We will only
consider this magnetically-charged case. We constrain the parameters so
that $G(\xi)$ has four roots, which we label by $\xi_1 \leq \xi_2 <
\xi_3 < \xi_4$. To obtain the right signature, we restrict $x$ to
$\xi_3 \leq x \leq \xi_4$, and $y$ to $-\infty < y \leq x$. The inner
black hole horizon lies at $y=\xi_1$, the outer black hole horizon at
$y=\xi_2$, and the acceleration horizon at $y=\xi_3$. The axis
$x=\xi_4$ points towards the other black hole, and the axis $x =
\xi_3$ points towards infinity. To avoid having a conical singularity
between the two black holes, we choose
\begin{equation} \label{dphi}
  \Delta \varphi = \frac{4\pi}{|G'(\xi_4)|},
\end{equation}
which implies that there will be a conical deficit along $x=\xi_3$,
with deficit angle
\begin{equation}
  \delta = 2\pi \left( 1 - \left| \frac{G'(\xi_3)}{G'(\xi_4)} \right|
\right).
\end{equation}
Physically, we imagine that this represents a cosmic string of mass per
unit length $\mu = \delta/8\pi$ along $x=\xi_3$. At large spatial
distances, that is, as $x, y \to \xi_3$, the $C$ metric
(\ref{Cmetric}) reduces to flat space with conical deficit $\delta$ in
accelerated coordinates.  If we converted to cylindrical coordinates
$(t,z,\rho,\varphi)$ on the flat space, the acceleration horizon would
correspond to the surface $z=0$.

We might wonder whether it is possible to replace the conical
singularity in the $C$ metric with a real cosmic string. There are two
potential problems: first of all, we have to be concerned about the
effect of the string stress-energy on the geometry in the neighbourhood
of the black hole horizon. However, it was shown in \cite{ruth} that
real vortices could pierce the black hole event horizon, so we will
assume that this does not prevent the replacement.

Secondly, we might worry about having a string end in a black hole.
If the strings are topologically unstable (that is, there are
monopoles present before the phase transition at which the strings
form), then we know that the strings can end at monopoles.  But away
from the event horizon, the field around a charged black hole is very
similar to that around a monopole. It therefore seems reasonable to
expect that a string can end in a black hole. It has been argued that
any cosmic string can end on a black hole, even if the string is
topologically stable \cite{ruth} (this argument is also given in
\cite{gary}). However, Preskill has remarked \cite{pres} that strings
which are potentially the boundaries of domain walls {\it cannot} end
on black holes, as the boundary of a boundary is zero (this category
includes topologically stable global strings). For strings which
cannot be the boundaries of domain walls, however, the argument of
\cite{ruth} applies (contrary to the statements in an earlier version
of this paper).

We can obtain the Euclidean section of the $C$ metric by setting $t =
i\tau$ in (\ref{Cmetric}). To make the Euclidean metric positive
definite, we need to restrict the range of $y$ to $\xi_2 \leq y \leq
\xi_3$. There are then potentially conical singularities at $y=\xi_2$
and $y=\xi_3$, which have to be eliminated.  We can avoid having a
conical singularity at
$y=\xi_3$ by taking $\tau$ to be periodic with period
\begin{equation}
  \Delta \tau =\beta = \frac{4\pi}{G'(\xi_3)}.
\end{equation}
If we assume that the black holes are extreme, that is, $\xi_1 =
\xi_2$, then the spatial distance from any other point to $y=\xi_2$ is
infinite, and so $\xi_2 < y \leq \xi_3$ on the Euclidean section, so
the conical singularity at $y=\xi_2$ is not part of the Euclidean
section.  Alternatively, if we assume $\xi_1 < \xi_2$, we can avoid
having a conical singularity at $y=\xi_2$ by taking the two horizons
to have the same temperature, so that both conical singularities can
be removed
by the same choice of $\Delta \tau$. This implies
\begin{equation}
  \xi_2 - \xi_1 = \xi_4 - \xi_3.
\end{equation}
As in the Ernst case, the former solution has topology $S^2 \times R^2
- \{pt\}$, while the latter has topology $S^2 \times S^2 - \{pt\}$.

We can obtain an instanton by slicing the Euclidean section in half
along a surface $\tau = 0, \beta/2$. This instanton will interpolate
between a slice of flat space with a conical deficit and a slice of
the $C$ metric, that is, a slice containing two black holes with
conical deficits running between the black holes and infinity. Thus,
this instanton can be used to model the breaking of a long piece of
cosmic string, with oppositely-charged black holes being created at
the free ends. If $\xi_2 = \xi_1$, the black holes are extreme, while
if $\xi_2 - \xi_1 = \xi_4 - \xi_3$, the black holes are non-extreme.

The semi-classical approximation to the amplitude for the string to
break (per unit length per unit time) will be given by $e^{-I_i}$, where
$I_i$ is the action of this instanton. Using the fact that the extrinsic
curvature of the slice $\tau=0, \beta/2$ vanishes, we can show that
the probability for the string to break is $e^{-I_E}$, where $I_E$ is now
the action of the whole Euclidean solution \cite{duality}.

We will calculate the action of the Euclidean section following the
technique used in \cite{haho,entar}. In fact, the calculation is very
similar to the calculation of the action in \cite{entar}. Since the
solution is static, the action can be written in the form
\begin{equation}
  I_E = \beta H - \frac{1}{4} \Delta {\cal A}
\end{equation}
in the extreme case, and
\begin{equation}
  I_E = \beta H - \frac{1}{4} (\Delta {\cal A}+ {\cal A}_{bh})
\end{equation}
in the non-extreme case, where the Hamiltonian is
\begin{equation}
  H = \int_\Sigma N {\cal H} - \frac{1}{8\pi} \int_{S^2_\infty} N (^2 K
  - ^2 K_0),
\end{equation}
$\Delta {\cal A}$ is the difference in area of the acceleration
horizon, ${\cal A}_{bh}$ is the area of the black hole event
horizon, $\Sigma$ is a surface of constant $\tau$, and $S^2_\infty$ is
its boundary at infinity.

Since the volume term in the Hamiltonian is proportional to the
constraint $\cal H$, which vanishes on solutions of the equations of
motion, the Hamiltonian is just given by the surface term. In the
surface term, $^2 K$ is the extrinsic curvature of the surface
embedded in the $C$ metric, while $^2 K_0$ is the extrinsic curvature
of the surface embedded in the background, flat space with a conical
deficit. We actually take a boundary `near infinity', and then take
the limit as it tends to infinity after calculating the Hamiltonian.
We choose the boundary in the $C$ metric to be at $x-y = \epsilon_c$.

We want to ensure that we take the same boundary in calculating the
two components of the Hamiltonian, which is achieved by requiring that
the intrinsic metric on the boundary as embedded in the two spacetimes
agree. We therefore want to write the flat background metric in a coordinate
system which makes it easy to compare it to the $C$ metric. We can in
fact write the flat metric as
\begin{eqnarray}  \label{flatacc}
ds^2 &=& \bar{A}^{-2} (x-y)^{-2} \left[ (1- y^2) dt^2 - (1- y^2)^{-1}
dy^2 \right. \\ && \left. +
(1-x^2)^{-1} dx^2 + (1-x^2) d\varphi^2 \right],
 \nonumber
\end{eqnarray}
where $\Delta \varphi = 2\pi -\delta$. Note that $\bar{A}$ represents
a freedom in the choice of coordinates, and $x$ is restricted to $-1
\leq x \leq 1$. A suitable background for the action calculation can
be obtained by taking $t=i\tau$ and $y \leq -1$ in (\ref{flatacc}). We
now take the boundary in the flat metric (\ref{flatacc}) to lie at
$x-y = \epsilon_f$.  It is easy to see that the induced metrics
on the boundary will agree if we take
\begin{equation}
  \bar{A}^2 = - \frac{G'(\xi_3)^2}{2 G''(\xi_3)} A^2, \epsilon_f = -
  \frac{G''(\xi_3)}{G'(\xi_3)} \epsilon_c.
  \label{epsrel}
\end{equation}

We can now calculate the two contributions to the Hamiltonian: the
contribution from the $C$ metric is (neglecting terms of order
$\epsilon_c$ and higher)
\begin{equation}
\int_{S^2_\infty} N {}^2 K= \frac{8\pi}{A^2 \epsilon_c |G'(\xi_4)|}
  \left[ 1 - \frac{1}{4} \epsilon_c \frac{G''(\xi_3)}{G'(\xi_3)} \right],
  \label{cmc}
\end{equation}
while the contribution from the flat background is
\begin{equation}
\int_{S^2_\infty} N {}^2 K_0 = \frac{4\pi}{\bar{A}^2 \epsilon_f} \left|
\frac{G'(\xi_3)}{G'(\xi_4)} \right| \left( 1 + \frac{1}{4} \epsilon_f \right).
  \label{fmc}
\end{equation}
Using (\ref{epsrel}), we see that these two surface terms are equal to
this order. Thus, in the limit $\epsilon \to 0$, the Hamiltonian
vanishes.

Thus, the action is just given by
\begin{equation}
  I_E = - \frac{1}{4} \Delta {\cal A}
\end{equation}
in the extreme case and
\begin{equation}
  I_E = - \frac{1}{4} (\Delta {\cal A}+ {\cal A}_{bh})
\end{equation}
in the non-extreme case. Note that, as in the Ernst case
\cite{entar}, the probability to produce a pair of extreme black holes
when the string breaks is suppressed relative to the probability to
produce a pair of non-extreme black holes by a factor of $e^{{\cal
A}_{bh}/4}$.

The area of the black hole horizon is
\begin{equation}
  {\cal A}_{bh} = \int_{y=\xi_2} \sqrt{g_{xx} g_{\varphi \varphi}} dx
  d\varphi = \frac{4\pi (\xi_4 - \xi_3)}{A^2 |G'(\xi_4)|(\xi_3- \xi_2)
    (\xi_4 - \xi_2)}.
\end{equation}
To calculate the difference in area of the acceleration horizon, we
calculate the area inside a circle at large radius in both the $C$
metric and the background, and take the difference. The area of the
acceleration horizon $y=\xi_2$ inside a circle at $x = \xi_3+
\epsilon_c$ in the
$C$ metric is
\begin{eqnarray}
  {\cal A}_c &=& \int_{y=\xi_3} \sqrt{g_{xx} g_{\varphi \varphi}} dx
  d\varphi \\ &=& - \frac{\Delta \varphi}{A^2 (\xi_4 - \xi_3)} +
  \frac{\Delta \varphi}{A^2 \epsilon_c} \nonumber \\ &=& -
  \frac{4\pi}{A^2 |G'(\xi_4)| (\xi_4-\xi_3)} + \pi \rho_c^2 \left|
  \frac{G'(\xi_3)}{G'(\xi_4)}\right|, \nonumber
\end{eqnarray}
where $\rho_c^2 = 4 /[ A^2 G'(\xi_3)\epsilon_c]$. The area of the
acceleration horizon $z=0$ inside a circle at $\rho=\rho_f$ in
the flat background is
\begin{equation}
  {\cal A}_f = \int \sqrt{g_{\rho\rho} g_{\varphi \varphi}} d\rho
  d\varphi = \pi \rho_f^2 \left| \frac{G'(\xi_3)}{G'(\xi_4)}\right|.
\end{equation}
To ensure that we are using the same boundary in calculating these two
components, we require that the proper length of the boundary be the
same. This gives
\begin{equation}
  \rho_f = \rho_c \left[ 1+ \frac{G''(\xi_3)}{G'(\xi_3)^2 A^2
    \rho_c^2} \right].
\end{equation}
We can now calculate the difference in area; it is
\begin{eqnarray}
  \Delta {\cal A} &=& {\cal A}_c - {\cal A}_f = - \frac{4\pi}{A^2
    |G'(\xi_4)| } \left[ \frac{1}{(\xi_4 - \xi_3)} +
  \frac{G''(\xi_3)}{2 G'(\xi_3)} \right] \\ &=& - \frac{4\pi}{A^2
    |G'(\xi_4)| } \left[ \frac{2}{(\xi_3 - \xi_1)} +
  \frac{(\xi_2-\xi_1) }{(\xi_3-\xi_2) (\xi_3 - \xi_1)} \right].
  \nonumber
\end{eqnarray}

In the extreme case, $\xi_2 = \xi_1$, so the action is
\begin{equation} \label{ac1}
  I_E = -\frac{1}{4} \Delta {\cal A} = \frac{2\pi}{A^2 |G'(\xi_4)|
    (\xi_3 - \xi_1)}.
\end{equation}
In the non-extreme case, the action is
\begin{equation} \label{ac2}
  I_E = -\frac{1}{4}( \Delta {\cal A}+ {\cal A}_{bh}) = \frac{2\pi}{A^2
    |G'(\xi_4)| (\xi_3 - \xi_1)},
\end{equation}
where we have used the condition $\xi_2 - \xi_1 = \xi_4 -\xi_3$ to
cancel the second contribution from $\Delta {\cal A}$ with the
contribution from ${\cal A}_{bh}$.

The limit $r_+ A \ll 1$ may be regarded as a point particle limit, as
it represents a black hole small on the scale set by the acceleration.
It is in this limit that we would expect to reproduce the result of
\cite{vil} on the probability for strings to break, forming monopoles
at the free ends.  In this limit, both the extreme and non-extreme
instantons satisfy $r_+ \approx r_-$ (that is, $q \approx m$). The mass
per unit length of the string in this limit is
\begin{equation}
  \mu \approx  r_+ A,
\end{equation}
and the action (\ref{ac1},\ref{ac2}) in this limit is
\begin{equation}
I_E \approx \frac{\pi r_+}{A} \approx \frac{\pi m^2}{\mu},
  \label{aclim}
\end{equation}
in agreement with the calculation of \cite{vil}, which found that the
action was $ I_E = \pi M_m^2/\mu$, where $M_m$ was the monopole mass.

If it is not topologically stable, the string is far more likely to
break and form monopoles than it is to break and form black holes, as
we do not expect that this semi-classical treatment is appropriate if
the black hole mass $m$ is less than the Planck mass, while the
monopole mass is typically of the order of $10^{-2} M_{Planck}$.
However, even certain kinds of strings that would be topologically
stable in flat space can break by the pair creation of black holes
\cite{gary,pres}. Since the mass per unit length $\mu$ for realistic
cosmic strings is typically of the order $10^{-6} M_{Planck} /
l_{Planck}$, breaking to form either monopoles or black holes is
extremely rare, and the effect of these tunnelling processes on cosmic
string dynamics is negligible.

\medskip {\large \bf Acknowledgements:} S.F.R. thanks the Association
of Commonwealth Universities and the Natural Sciences and Engineering
Research Council of Canada for financial support. We acknowledge
helpful conversations with Rob Caldwell, and thank Ruth Gregory for
giving a talk which inspired this work. We also thank John Preskill
and Gary Horowitz for pointing out our error in the discussion of the
breaking of topologically stable strings.

\end{document}